\documentclass{PoS}
\usepackage{natbib}

\title{Performance of the MAGIC telescopes after the major upgrade}

\ShortTitle{Performance of the MAGIC telescopes after the major upgrade}

\author{\speaker{Julian Sitarek}\\
        University of \L\'od\'z, PL-90236 Lodz, Poland, \\
        IFAE, Campus UAB, E-08193 Bellaterra, Spain
        E-mail: \email{jsitarek@uni.lodz.pl}}

\author{Emiliano Carmona\\
        E-mail: \email{emilianocarm@gmail.com}}

\author{Pierre Colin\\
        Max-Planck-Institut f\"ur Physik, D-80805 M\"unchen, Germany
        E-mail: \email{colin@mppmu.mpg.de}}

\author{Daniel Mazin\\
        ICRR, The University of Tokyo
        E-mail: \email{mazin@icrr.u-tokyo.ac.jp}}

\author{Diego Tescaro\\
        Inst. de Astrof\'isica de Canarias, E-38200 La Laguna, Tenerife, Spain
        E-mail: \email{diego.tescaro@gmail.com}}

\author{for the MAGIC collaboration}

\abstract{
MAGIC is a system of two Imaging Atmospheric Cherenkov Telescopes located on the Canary island of La Palma, Spain. 
During summer 2011 and 2012 it underwent a major upgrade. 
The main subsystems upgraded were the MAGIC-I camera and its trigger system and the readout system of both telescopes. 
We use observations of the Crab Nebula taken at low and medium zenith angles to assess the key performance parameters of the MAGIC stereo system. 
For low zenith angle observations, the standard trigger threshold of the MAGIC telescopes is about 50 GeV. 
The integral sensitivity for point-like sources with Crab Nebula-like spectra above 220\,GeV is $(0.66\pm0.03)\%$ of Crab Nebula flux in 50\,h of observations. 
The angular resolution, defined as the sigma of a 2-dimensional Gaussian distribution, at energies of a few hundred GeV is below $0.07^\circ$, while the energy resolution is around 16\%. 
We investigate the effect of the systematic uncertainty on the data taken with the MAGIC telescopes after the upgrade. 
We estimate that the systematic uncertainties can be divided in the following components: $< 15\%$ in energy scale, $11-18\%$ in flux normalization and $\pm0.15$ for the slope of the energy spectrum.
}

\FullConference{The 34th International Cosmic Ray Conference,\\
		30 July- 6 August, 2015\\
		The Hague, The Netherlands}

\begin{document}
\section{Introduction}
MAGIC (Major Atmospheric Gamma Imaging Cherenkov telescopes) is a system of two 17\,m diameter Imaging Atmospheric Cherenkov Telescopes (IACTs). 
The telescopes are located at a height of 2200 m a.s.l. on the Roque de los Muchachos Observatory on the Canary Island of La Palma, Spain ($28^\circ$N, $18^\circ$W).
They are used for observations of particle showers produced in the atmosphere by very high energy (VHE, $\gtrsim30\,$GeV) $\gamma$-rays. 
Between summer 2011 and 2012 the telescopes went through a major upgrade  \citep{al15a}, carried out in two stages. 
In the first part of the upgrade, in summer 2011, the readout systems of both telescopes were upgraded \citep{magic_drs4}.
The multiplexed FADCs used before in MAGIC-I \citep{magic_mux} as well as the Domino Ring Sampler version 2 used in MAGIC-II (DRS2; \citealp{magic_daq}) have been replaced by  Domino Ring Sampler version 4 chips (DRS4; \citealp{ritt_drs4}).
The switch to DRS4 based readout allowed to eliminate the $\sim 10\%$ dead time present in the previous system due to the DRS2 chip and lowered the electronic noise of the readout.  
In summer 2012 the second stage of the upgrade followed with an exchange of the camera of the MAGIC-I telescope to a uniformly pixelized one \citep{al15a}. 
Currently each MAGIC camera is equipped with 1039 photomultipliers (PMTs) and has a total field of view of $\sim3.5^\circ$.
The upgrade of the camera allowed to increase the area of the trigger region in MAGIC-I by a factor of 1.7 to the value of $4.8$ square degrees. 

\section{Data sample and analysis}
We evaluate the performance of the MAGIC telescopes using  data from the Crab Nebula, the so-called standard candle of VHE $\gamma$-ray astronomy.
The data have been taken between October 2013 and January 2014 in the so-called wobble mode \citep{fo94}, i.e. with the source position offset by $0.4^\circ$ from the center of the camera. 
This method allows to simultaneously estimate the background from reflected positions in the sky at the same offset. 
The data are divided into two sub-ranges according to the zenith angle of the observations: $5-30^\circ$ (11\,h) and $30-45^\circ$ (4\,h).
The data have been analyzed using the standard MAGIC tools: MARS (MAGIC Analysis and Reconstruction Software, \citealp{magic_mars}).
The analysis chain (extraction of Cherenkov light pulses, calibration to photoelectrons (phe), image cleaning and parametrization, $\gamma$/hadron separation, estimation of energy and arrival direction) is described in detail in \cite{al15b}.

\section{Performance parameters}
Here we present the most important performance parameters of the MAGIC telescopes: energy threshold, angular resolution, energy resolution and sensitivity. 

\subsection{Energy threshold}
The energy resolution of an IACT is normally defined as the peak of the differential energy distribution obtained with Monte Carlo (MC) events weighted to a particular spectral shape. 
In here we assume for the calculation the commonly used spectral index of $-2.6$. 

\begin{figure}[t]
\centering 
\includegraphics[width=0.49\textwidth]{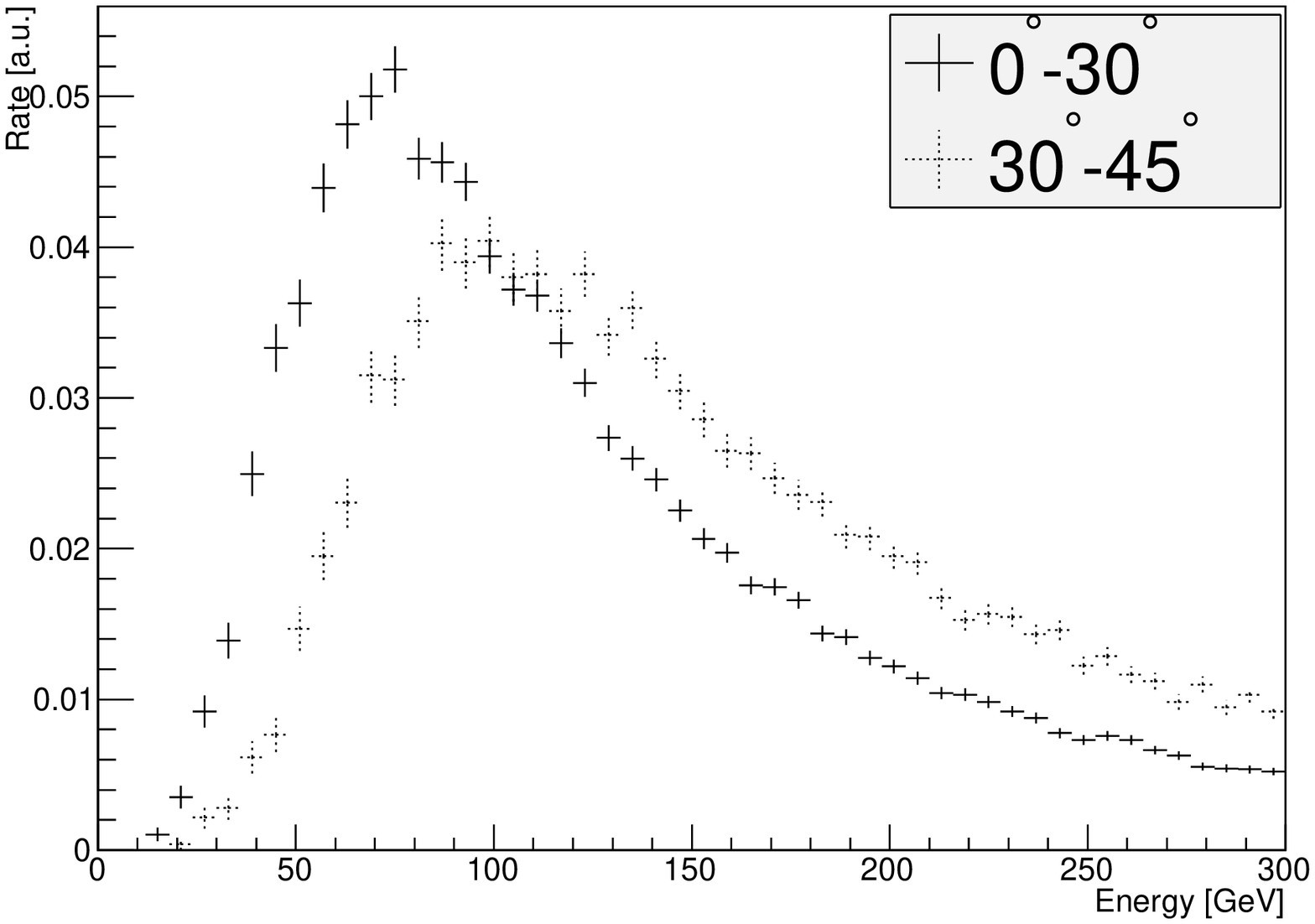}
\includegraphics[width=0.49\textwidth]{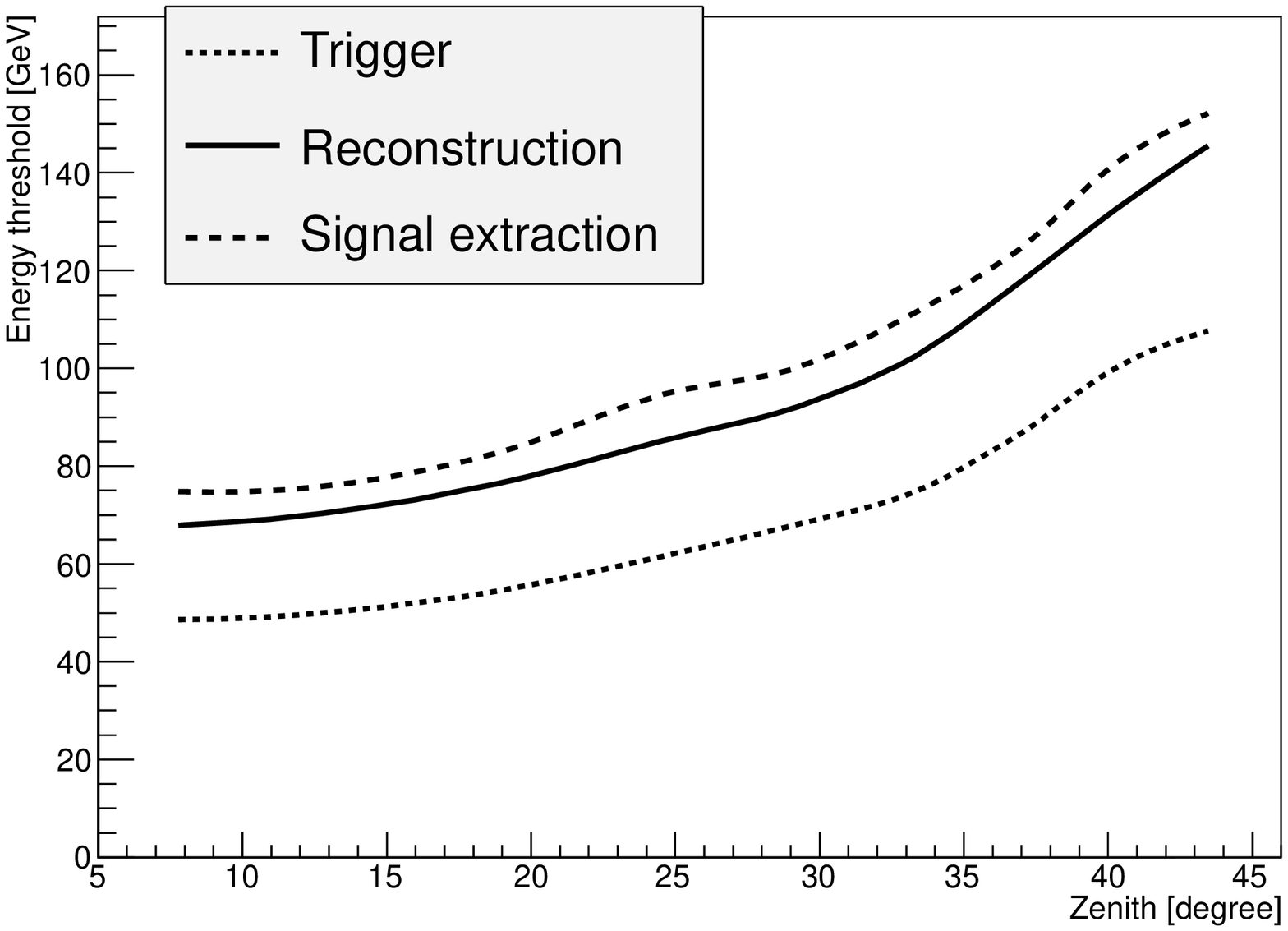}
\caption{
Left panel: rate of MC $\gamma-$ray events (in arbitrary units) surviving the image cleaning with at least 50 phe for a source with a spectral index of $-2.6$ in the zenith range  $<30^\circ$ (solid line) and $30-45^\circ$ (dotted line). 
Right panel: threshold of the MAGIC telescopes as a function of the zenith angle of the observations at the trigger level (the dotted curve), only events with images that survived image cleaning in each telescope with at least 50\,phe (solid line) and with additional cuts of $Hadronness<0.5$ and $\theta^2<{0.03^\circ}^2$ (dashed line). Both panels reproduced from \cite{al15b}.
}\label{fig:threshold}
\end{figure}

In the left panel of Fig.~\ref{fig:threshold} we show the differential rate plot in two zenith angle ranges for events that survived image reconstruction in both telescopes.
For low zenith angle, i.e. $<30^\circ$, the reconstruction threshold energy is $\sim 70\,$GeV.
Note however that the peak is broad and extends far to lower energies, allowing to evaluate the performance of the telescopes and to obtain scientific results below such defined threshold.
In the right panel of Fig.~\ref{fig:threshold} we show how the energy threshold of the MAGIC telescopes depends on the zenith angle of observations.
The threshold is quite stable for low zenith angle observations.
It increases rapidly for higher values of zenith angles, due to both the larger absorption of the Cherenkov light in the atmosphere and the dilution of the photons reaching the ground over a larger light pool.

\subsection{Energy resolution}
\begin{figure}[tp]
\centering 
\includegraphics[width=0.49\textwidth]{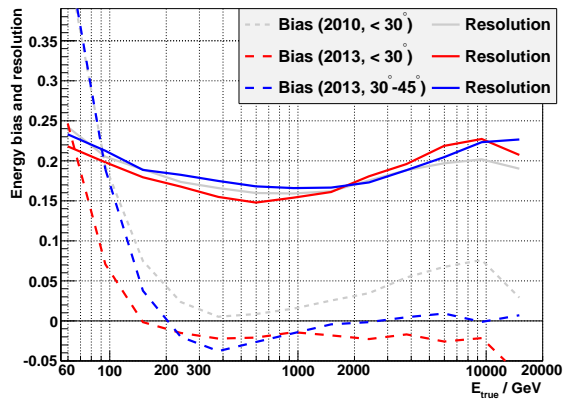}
\caption{
Energy resolution (solid lines) and bias (dashed lines) obtained from the MC simulations of $\gamma$ rays at low (red) and medium (blue) zenith angles.
Events have been weighted in order to represent a spectrum with a slope of $-2.6$. 
For comparison, pre-upgrade values from \cite{magic_stereo} are shown in gray lines.
The figure is reproduced from \cite{al15b}.
}\label{fig:erec}
\end{figure}

We evaluate the performance of the energy reconstruction with $\gamma-$ray MC simulations. 
The simulations are divided into bins of true energy (5 bins per decade). 
In each bin we construct a distribution of $(E_{est}-E_{true})/E_{true}$ and fit it with a Gaussian function. 
The Gaussian function describes well the distribution in the central region, but not at the edges (see \citealp{al15b}). 
The energy resolution and bias are defined as the standard deviation and the mean value respectively of this fit.
Both quantities are plotted as a function of the true energy of the $\gamma$ rays in Fig.~\ref{fig:erec}.

For low zenith angle observations in the energy range of a few hundred GeV the energy resolution is as low as $\sim15\%$. 
For higher energies it degrades due to an increasing fraction of truncated images, and showers with high impact parameters as well as worse statistics in the training sample.  
At low energies the energy resolution is degraded, due to worse precision in the image reconstruction (in particular the impact parameters), and higher internal relative fluctuations of the shower. 
Above a few hundred GeV the absolute value of the bias is below a few percent. 
At low energies ($\lesssim100\,$GeV) the estimated energy bias rapidly increases due to the threshold effect. 
For observations at higher zenith angles the energy resolution is similar.
Since an event of the same energy observed at higher zenith angle will produce a smaller image, the energy resolution at the lowest energies is slightly worse. 
On the other hand, at multi-TeV energies, the showers observed at low zenith angle can be partially truncated at the edge of the camera or saturate some of the pixels.
Therefore the energy resolution at those energies is slightly better for higher zenith angle observations. 
On the other hand, as the energy threshold shifts with increasing zenith angle, the energy bias at energies below 100 GeV is much stronger for higher zenith angle observations. 

\subsection{Angular resolution}\label{sec:angres}
Following the approach in \cite{magic_stereo}, we investigate the angular resolution of the MAGIC telescopes using two commonly used methods. 
In the first approach we define the angular resolution $\Theta_{\rm Gaus}$ as the standard deviation of a 2-dimensional Gaussian fitted to the distribution of the reconstructed event directions of the $\gamma$-ray excess. 
Such a 2-dimensional Gaussian in the $\theta_x$ and $\theta_y$ space will correspond to an exponential fitting function for $\theta^2$ distribution.
The fit is performed in a narrow range, $\theta^2<0.025{^\circ}^2$.
Therefore it is an appropriate quantity for estimating performance of the instrument in case of a search for small extensions (comparable with angular resolution) in VHE $\gamma-$ray sources. 
In the second method we compute an angular distance, $\Theta_{0.68}$, around the source, which contains 68\% of the excess events.
This method is more sensitive to long tails in the distribution of reconstructed directions. 
Note that while both numbers assess the angular resolution of the MAGIC telescopes, their absolute values are different.
For a purely Gaussian distribution $\Theta_{\rm Gaus}$ would correspond to only 39\% containment radius of $\gamma$-rays originating from a point-like source and $\Theta_{0.68}\approx1.5\,\Theta_{\rm Gaus}$.

\begin{figure*}[pt]
\centering 
\includegraphics[width=0.49\textwidth]{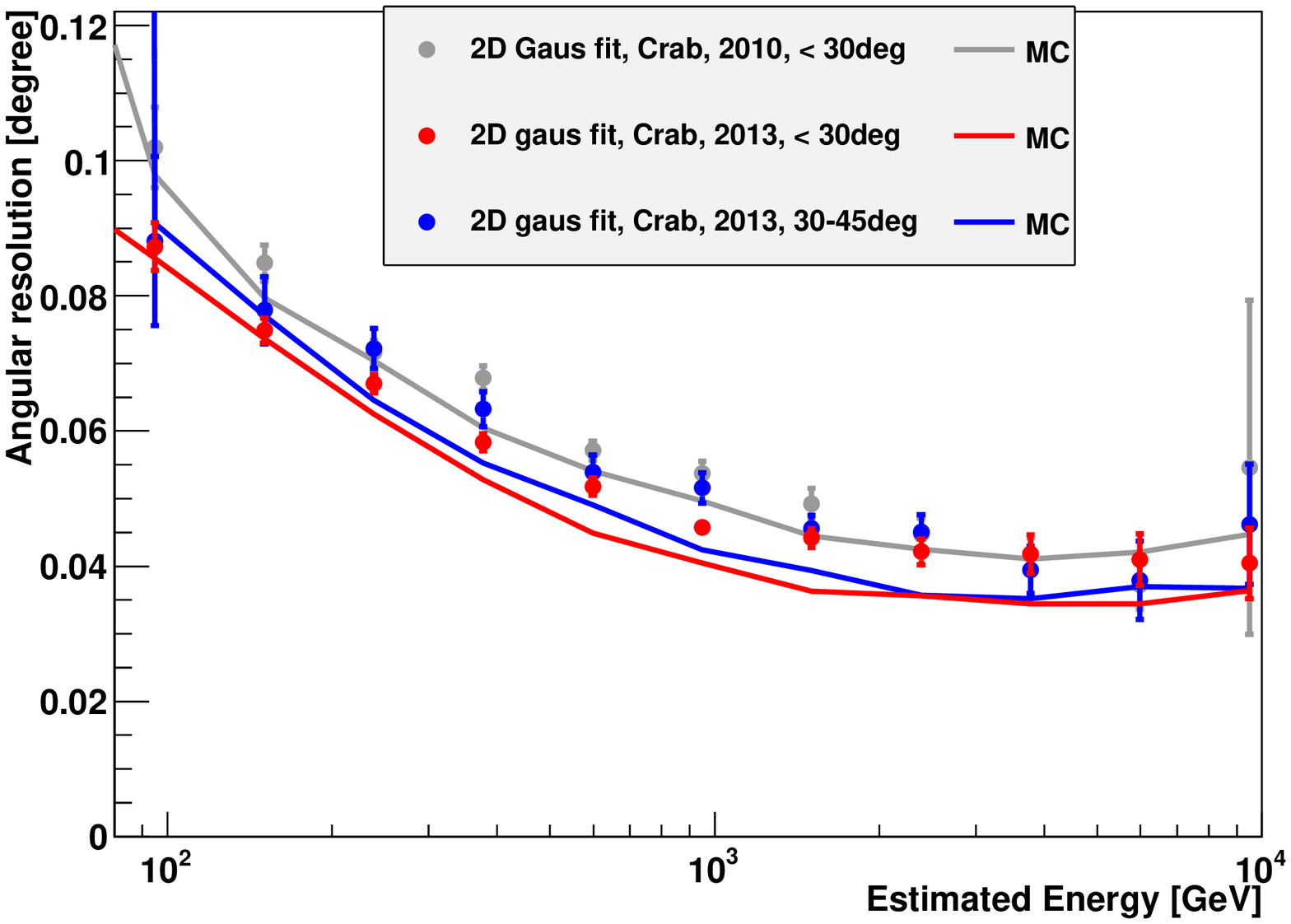}
\includegraphics[width=0.49\textwidth]{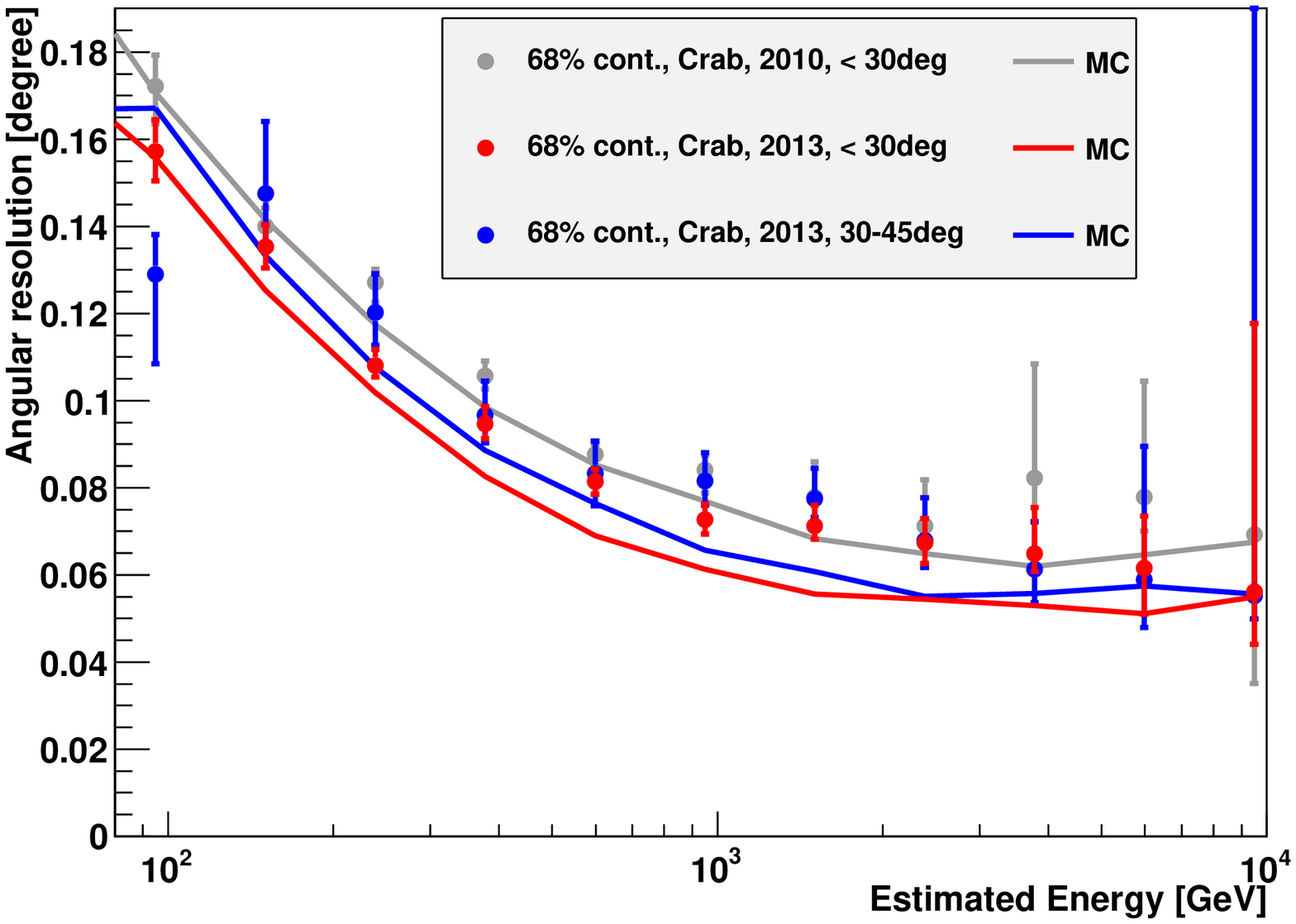}
\caption{
Angular resolution of the MAGIC telescopes after the upgrade as a function of the estimated energy obtained with the Crab Nebula data sample (points) and MC simulations (solid lines). 
Left panel: 2D Gaussian fit, right panel: 68\% containment radius.
Red points: low zenith angle sample, blue points: medium zenith angle sample.
For comparison the low zenith angle pre-upgrade angular resolution is shown as gray points \cite{magic_stereo}.
Reproduced from \cite{al15b}.
}\label{fig:angres}
\end{figure*}

We use the low and medium zenith angle samples of the Crab Nebula to investigate the angular resolution.
%
The angular resolution obtained with both methods is shown in Fig.~\ref{fig:angres}.
At $250\,$GeV the angular resolution (computed with the 2D-Gaussian fit) is $0.07^\circ$.
It improves with energy, as larger images are better reconstructed, reaching a plateau of $\sim 0.04^\circ$ above a few TeV. 
The angular resolution improved by about 5-10\% after the upgrade.
Due to the improvement in angular resolution the small difference between the angular resolution obtained with MC simulations and the Crab Nebula data, also present in the pre-upgrade data, is slightly more pronounced.
The difference of $\sim 10-15\%$ is visible at higher energies and corresponds to an additional $0.02^\circ$ systematic random component (i.e. added in quadrature) between MC and data.
Study of the reconstructed source direction spread in each of 10 nightly sub-samples of the Crab Nebula data suggest a similar, $\lesssim 0.02^\circ$, systematic uncertainty on the pointing position \citep{al15b}.

\subsection{Sensitivity}\label{sec:sens}
For a weak source, the significance of an excess of $N_{\rm excess}$ events over a perfectly-well known background of $N_{\rm bkg}$ events can be calculated as $N_{\rm excess}/\sqrt{N_{\rm bkg}}$.
Therefore, one defines the sensitivity $S_{\rm Nex/\!\sqrt{Nbkg}}$ as the flux of a source giving $N_{\rm excess}/\sqrt{N_{\rm bkg}}=5$ after 50$\,$h of effective observation time.
For a more realistic estimation of the sensitivity, we apply two additional conditions: $N_{\rm excess}>10$ and $N_{\rm excess}> 0.05 N_{\rm bkg}$.
The first condition assures that the Poissonian statistics of the number of events can be approximated by a Gaussian distribution.
The second one counteracts the effect of small systematic discrepancies between the ON and OFF distributions, which could mimic a statistically significant signal if the residual background rate is large. 

\begin{figure}[t!]
\centering 
\includegraphics[width=0.49\textwidth]{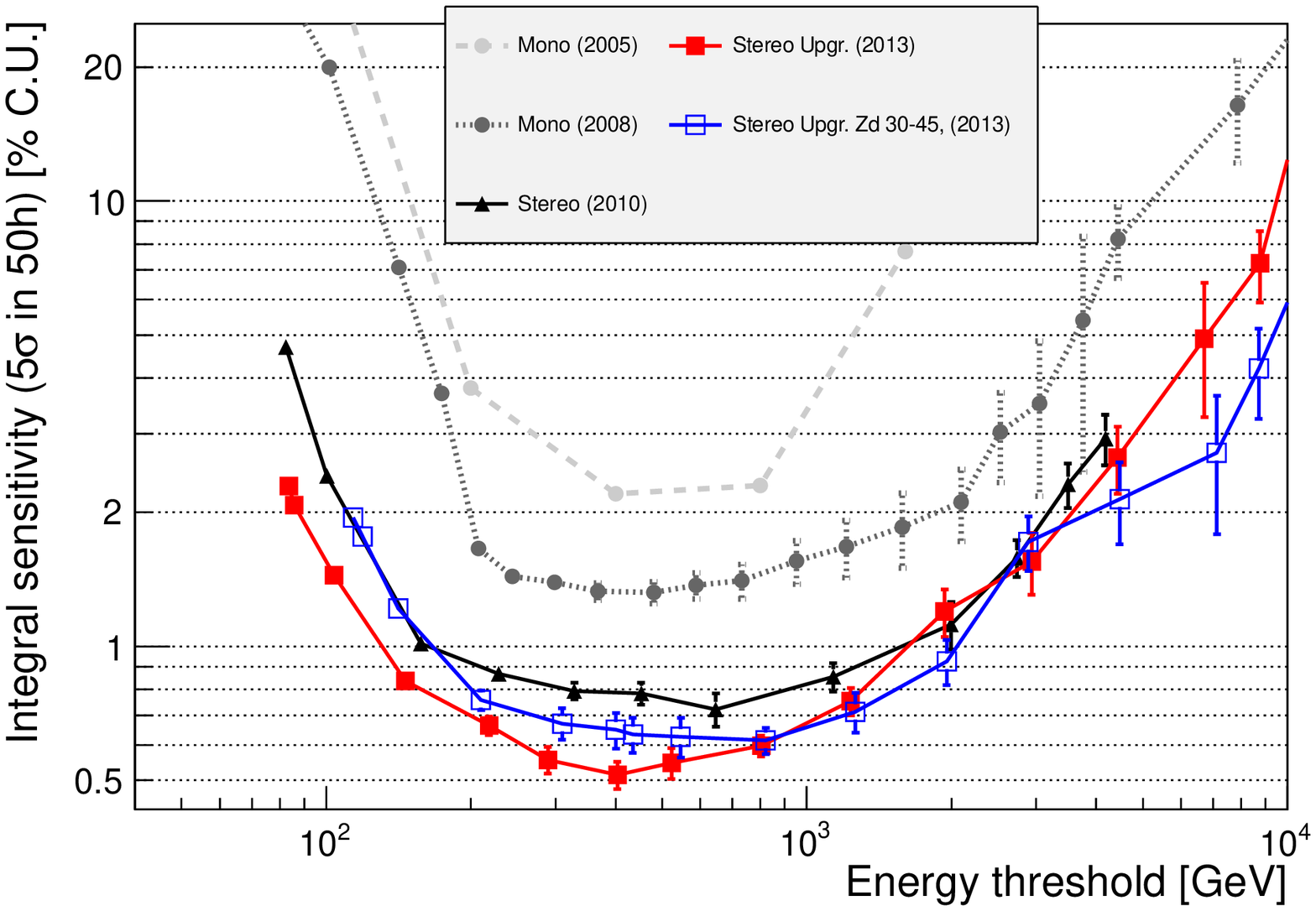}
\includegraphics[width=0.49\textwidth]{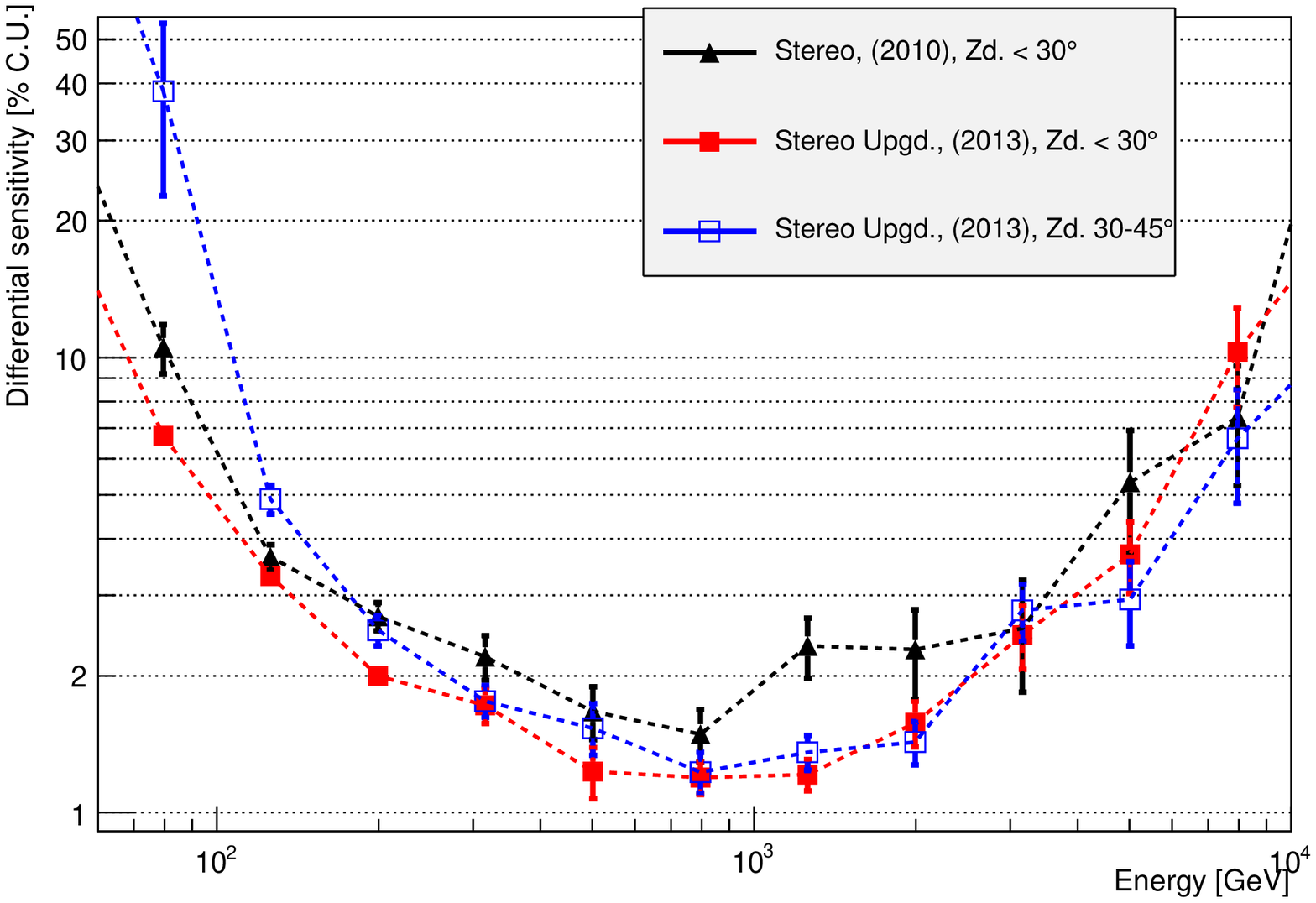}
\caption{
Integral (left panel) and differential (right panel, 5 bins per decade in energy) sensitivity of the MAGIC telescopes expressed in Crab Units. 
The sensitivity is defined as the flux of the source in a given energy range for which $N_{\rm excess}/\sqrt{N_{\rm bkg}}=5$ and $N_{\rm excess}>10$, $N_{\rm excess}> 0.05 N_{\rm bkg}$ after 50$\,$h of effective time.
Left panel:
Gray circles: sensitivity of the MAGIC-I single telescope with the Siegen (light gray, long dashed, \citet{magic_crab}) and MUX readouts (dark gray, short dashed, \citet{magic_stereo}).
Black triangles: stereo before the upgrade \citep{magic_stereo}.
Squares: stereo after the upgrade: zenith angle below $30^\circ$ (red, filled), $30-45^\circ$ (blue, empty).
Both panels are reproduced from \cite{al15b}.
}\label{fig:sens}
\end{figure}

The integral sensitivity of the different phases of the MAGIC experiment for a source with a Crab Nebula-like spectrum are shown in the left panel of Fig.~\ref{fig:sens}. 
In order to find the optimal cut values in $Hadronness$  ($\gamma$/hadron separation parameter) and $\theta^2$ in an unbiased way, we split the total sample of Crab Nebula data into two independent sub-samples (\emph{training} and \emph{test}) of similar size and similar observation conditions. 
Different energy thresholds are achieved by varying a cut in the total number of photoelectrons of the images (for points $<300\,$GeV) or in the estimated energy of the events (above 300\,GeV).
For each energy threshold we perform a scan of cuts on the \emph{training} sub-sample.
The pair of cuts which provides the best sensitivity on the \emph{training} sub-sample is selected and applied to \emph{test} sample to obtain the final sensitivity value.
The threshold itself is estimated as the peak of true energy distribution of MC events with a $-2.6$ spectral slope to which the same cuts were applied.

The integral sensitivity evaluated above is valid only for sources with a Crab Nebula-like spectrum. 
To assess the performance of the MAGIC telescopes for sources with an arbitrary spectral shape, we compute the differential sensitivity, i.e. the sensitivity in narrow bins of energy (5 bins per decade).
The differential sensitivity is plotted for low and medium zenith angles in right panel of Fig.~\ref{fig:sens}.
The sensitivity computed also according to different definitions is presented in \cite{al15b}.

The upgrade of the MAGIC-I camera and readout of the MAGIC telescopes has lead to a significant improvement in sensitivity over the whole investigated energy range.
The integral sensitivity reaches down to about 0.55\% of C.U.\footnote{Crab Unit, i.e. the flux of Crab Nebula, which is a standard candle in VHE $\gamma$-ray astronomy.} around a few hundred GeV in 50\,h of observations.
The improvement in the performance is especially evident at the lowest energies. 
In particular, in the energy bin 60-100 GeV, the differential sensitivity decreased from 10.5\% C.U. to 6.7\% C.U. reducing the needed observation time by a factor of 2.5.
Observations at medium zenith angle have naturally higher energy threshold, degrading the performance at the lowest energies. 
On the other hand such observations provide a small boost in the sensitivity at the TeV energies. 

\section{Systematic uncertainties}
The systematic uncertainties of the IACT technique are a combination of multiple small individual factors which are only known with limited precision, and possibly evolve in time. 
Most of those factors (e.g. uncertainties connected with the atmosphere, reflectivity of the mirrors, etc.) were not affected by the upgrade. 
On the other hand the upgrade led to a decrease of the systematic uncertainty connected to background homogeneity \citep{al15b}.
The total systematic uncertainty can be divided into 3 components: pure flux normalization uncertainty, uncertainty in the energy scale (which can be converted into an additional uncertainty in the flux normalization) and uncertainty in the spectral index.
The final systematic uncertainties have been reevaluated in \citet{al15b}.
The uncertainty on the flux normalization is  18\% at low energies ($\lesssim100\,$GeV) and 11\% in the energy range of a few hundred GeV.
At the highest energies, $\gtrsim 1\,$TeV, due to more pronounced MC/data mismatches the systematic uncertainty is a bit higher, namely 16\%.
The uncertainty of the absolute energy scale of the MAGIC telescopes is below 15\%.
The systematic uncertainty on the reconstructed spectral slope of the sources is $\pm0.15$ for observations not strongly dominated by background events. 

\section{Conclusions}\label{sec:concl}
The upgrade of the readout and one of the cameras of the MAGIC telescopes have significantly improved their performance. 
The trigger threshold for low zenith angle observations is $\sim 50\,$GeV. 
For a Crab-like source, the best performance of the MAGIC telescopes is achieved at energies of a few hundred GeV.
At those energies the images are sufficiently large to provide enough information for efficient reconstruction, while the rapidly falling power-law spectrum of the Crab Nebula still provides enough statistics.
The energy resolution at these medium energies is as good as 16\% with a negligible bias and the angular resolution is $\lesssim 0.07^\circ$. 
The achieved sensitivity above 220\,GeV is $(0.66\pm0.03)\%$ of C.U. for 50\,h of observations. 
At the lowest energies, below $100\,$GeV, the performance has improved drastically comparing to the one presented in \citet{magic_stereo}, reducing the needed observation time by a factor of 2.5.

The performance of the MAGIC telescopes at medium zenith angles, $30-45^\circ$, is mostly similar to the one at low zenith angles. 
The higher threshold, however, significantly degraded all the performance parameters below $\sim200$\,GeV.
However, for the energies above a few TeV, better performance is achieved with observations at medium zenith angles.

Thanks to the improvement in the performance achieved after the upgrade, the MAGIC telescopes have reached an unprecedented sensitivity, which caused a significant boost in the scientific output of the experiment. 
In particular the excellent low energy performance allowed to expand the visible VHE $\gamma$-ray sky up to redshift $\sim0.94$ with the detection of two distant blazars: QSO B0218+357 \citep{atel6349} and PKS1441+25 \citep{atel7416}.

\section{Acknowledgments}

We would like to thank
the Instituto de Astrof\'{\i}sica de Canarias
for the excellent working conditions
at the Observatorio del Roque de los Muchachos in La Palma.
The financial support of the German BMBF and MPG,
the Italian INFN and INAF,
the Swiss National Fund SNF,
the ERDF under the Spanish MINECO (FPA2012-39502), and
the Japanese JSPS and MEXT
is gratefully acknowledged.
This work was also supported
by the Centro de Excelencia Severo Ochoa SEV-2012-0234, CPAN CSD2007-00042, and MultiDark CSD2009-00064 projects of the Spanish Consolider-Ingenio 2010 programme,
by grant 268740 of the Academy of Finland,
by the Croatian Science Foundation (HrZZ) Project 09/176 and the University of Rijeka Project 13.12.1.3.02,
by the DFG Collaborative Research Centers SFB823/C4 and SFB876/C3,
and by the Polish MNiSzW grant 745/N-HESS-MAGIC/2010/0.
JS is supported by Fundacja U\L .

\end{document}